\PassOptionsToPackage{draft}{hyperref} 
\documentclass[useAMS,usenatbib]{mnras}

\usepackage{graphicx}
\usepackage{verbatim}
\usepackage{bm}
\usepackage{pgf}
\usepackage{mathptmx}
\usepackage{amsmath,amssymb}
\usepackage{hyperref}
\usepackage{adjustbox}

\usepackage{ulem}

\title[Interstellar delays of J0613-0200 with LEAP]{Measuring Interstellar Delays of \psr over 7 years, using the Large European Array for Pulsars}
\author[Robert Main et al.]{R. A. Main$^{1}$\thanks{E-mail:ramain@mpifr-bonn.mpg.de}, S. A. Sanidas$^{2,3}$, J. Antoniadis$^{1,4,5}$, C. Bassa$^{6}$, S. Chen$^{7,8,9}$,  I. Cognard$^{9,7}$,\newauthor M. Gaikwad$^{1}$, H. Hu$^{1}$, G. H. Janssen$^{6,10}$, R. Karuppusamy$^{1}$,  M. Kramer$^{1,2}$, K. J. Lee$^{11}$, \newauthor K. Liu$^{1}$, G. Mall$^{1}$, J. W. McKee$^{12}$, M. B. Mickaliger$^{2}$,  D. Perrodin$^{13}$, B. W. Stappers$^{2}$,  \newauthor C. Tiburzi$^{6}$, O. Wucknitz$^{1}$, L. Wang$^{2, 14}$, W. W. Zhu$^{14}$ \\
$^{1}$Max-Planck-Institut f\"{u}r Radioastronomie, Auf dem H\"{u}gel 69, 53121, Bonn, Germany \\
$^{2}$Jodrell Bank Centre for Astrophysics, School of Physics and Astronomy, The University of Manchester, Manchester M13 9PL,UK \\
$^{3}$Anton Pannekoek Institute for Astronomy, University of Amsterdam, Science Park 904, 1098 XH Amsterdam, The Netherlands \\ 
$^{4}$Argelander Institut f\"{u}r Astronomie, Auf dem H\"{u}gel 71, 53121, Bonn, Germany\\
$^{5}$Institute of Astrophysics, FORTH, Dept. of Physics, University of Crete, Voutes, University Campus, GR-71003 Heraklion, Greece \\
$^{6}$ASTRON, the Netherlands Institute for Radio Astronomy, Oude Hoogeveensedijk 4, 7991 PD Dwingeloo, The Netherlands \\
$^{7}$Station de radioastronomie de Nan{\c c}ay, Observatoire de Paris, PSL Research University, CNRS/INSU F-18330 Nan{\c c}ay, France \\
$^{8}$FEMTO-ST, Department of Time and Frequency, UBFC and CNRS, F-25030 Besan\c{c}on, France \\
$^{9}$Laboratoire de Physique et Chimie de l'Environnement et de l'Espace LPC2E CNRS-Universit{\'e} d'Orl{\'e}ans, F-45071, Orl{\'e}ans, France \\
$^{10}$Department of Astrophysics/IMAPP, Radboud University, P.O. Box 9010, 6500 GL Nijmegen, The Netherlands \\
$^{11}$Kavli institute for astronomy and astrophysics, Peking University, Beijing 100871,P.R.China \\
$^{12}$Canadian Institute for Theoretical Astrophysics, University of Toronto, 60 St. George Street, Toronto, ON M5S 3H8, Canada \\
$^{13}$INAF - Osservatorio Astronomico di Cagliari, via della Scienza 5, I-09047 Selargius (CA), Italy \\
$^{14}$National Astronomical Observatories, Chinese Academy of Sciences, A20 Datun Rd, Chaoyang District, Beijing 100012, P.\,R.\,China \\
}

\begin{document}

\newcommand{\RM}[1]{\textcolor{purple}{RM: #1}}
\newcommand{\TS}[1]{\textcolor{olive}{TS: #1}}
\newcommand{\OW}[1]{\textcolor{blue}{[OW: #1]}}
\newcommand{\psr}{PSR J0613$-$0200 }
\newcommand{\dmu}{\,pc\,cm$^{-3}$ }
\newcommand{\new}[1]{\textcolor{purple}{#1}}

\newcommand{\mean}[1]{\left< #1 \right>}
\newcommand{\sub}[1]{_\text{#1}}

\date{Accepted . Received ; in original form}

\pagerange{\pageref{firstpage}--\pageref{lastpage}} 
\pubyear{2020}

\maketitle

\label{firstpage}


\begin{abstract}

Using data from the Large European Array for Pulsars (LEAP), and the Effelsberg telescope, we study the scintillation parameters of the millisecond pulsar \psr over a 7 year timespan.
The ``secondary spectrum" -- the 2D power spectrum of scintillation -- presents the scattered power as a function of time delay, and contains the relative velocities of the pulsar, observer, and scattering material.
We detect a persistent parabolic scintillation arc, suggesting scattering is dominated by a thin, anisotropic region. 
  
The scattering is poorly described by a simple exponential tail, with excess power at high delays; we measure significant, detectable scattered power at times out to $ \sim 5 \mu s$, and measure the bulk scattering delay to be between 
50 to 200\,ns
with particularly strong scattering throughout 2013. 
These delays are too small to detect a change of the pulse profile shape, yet they would change the times-of-arrival as measured through pulsar timing.
The arc curvature varies annually, and is well fit by a one-dimensional scattering screen $\sim 40\%$ of the way towards the pulsar, with a changing orientation during the increased scattering in 2013.  
Effects of uncorrected scattering will introduce time delays correlated over time in individual pulsars, and may need to be considered in gravitational wave analyses.
Pulsar timing programs would benefit from simultaneously recording in a way that scintillation can be resolved, in order to monitor the variable time delays caused by multipath propagation.

\end{abstract}


\section{Introduction}

Radio emission from pulsars experiences several propagation effects from the ionised interstellar medium (ISM), as the index of refraction varies with electron density and frequency.  The signal acquires a group delay $\Delta t$, known as dispersion, scaling as $\Delta t \propto \mathrm{DM}\,\nu^{-2}$, where DM is the integrated column density of free electrons, and $\nu$ is the observing frequency. 
Spatial variations in the electron density result in multi-path propagation, with deflected paths acquiring a geometric time delay from the path-length difference compared to the direct line-of-sight. When these delays are large (compared to the pulse duration), it is observed as scattering, a one-sided broadening of pulses often resembling an exponential tail.  When these delays are small, we observe it as scintillation, the constructive and destructive interference of different deflected images at the observer, resulting in a time and frequency dependence of the observed flux.  These delays are steeper in frequency than dispersion, and are expected to scale roughly as $\tau \sim \nu^{-4}$.

One of the central goals of pulsar timing is to directly detect gravitational waves, in a so-called pulsar timing array (\citealt{Hobbs2013}, \citealt{desvignes+16}, \citealt{verbiest+16}, \citealt{arzoumanian+18}).  The most stable pulsars are observed on weekly to monthly cadence over many years, and $\sim$nHz gravitational waves could be observed in timing residuals correlated in time and position on the sky \citep{hellings+83}.  This effect is expected to be tiny, with a fractional change of the arrival time compared to the gravitational wavelength of order $10^{-15}$, so it requires careful understanding of all other effects which would change the arrival times of pulses.  While PTA pulsars are selected for their stability, they all experience variable dispersion and scattering to some degree due to the relative motion of the pulsar and observer with respect to the ISM.  Variable dispersion measures have been measured and corrected using multifrequency data (eg. \citealt{keith+13}), while changes in scattering time are often estimated using the statistical relation between the scintillation bandwidth (the frequency width of scintillation) and scattering time (eg. \citealt{levin+16, shapiro+20}, see \citealt{verbiest+18} for a review of how these effects limit precision pulsar timing). Dispersion and scattering both scale strongly with frequency, and are often covariant. One can look for variable delays following a $\nu^{-2}$ and $\nu^{-4}$ scaling \citep{lam+19}; the technique of wide-band template matching has recently been developed as a way to jointly fit for these effects \citep{liu+14, pennucci+14, pennuci19, nanograv20b}.

In this paper, we begin to apply the methods of \citet{hemberger+08} to PTA pulsars, in which scintillation arcs are used to estimate time delays from multi-path propagation.
We analyse \psr over 7 years, in roughly monthly cadence, using data from the Large European Array for Pulsars (LEAP) \citep{stappers+11, bassa+16}, and a 3-month bi-weekly observing campaign using the 100--m Effelsberg radio telescope.  This pulsar is of particular interest; it shows the strongest evidence of a 15\,nHz strain, but since the signal appears most strongly in this pulsar, it is believed to arise from an unmodelled non-GW signal \citep{aggarwal+19}. In Section \ref{sec:theory}, we give an overview of some necessary background of scintillation, and summarise the methods of \citet{hemberger+08}.  In Section \ref{sec:data}, we describe our observations with the LEAP telescope, our short-term observing campaign with the Effelsberg telescope.  In Section \ref{sec:methods} we outline our methods, in Section \ref{sec:results} we present our results, and we discuss the ramifications and future prospects in Section \ref{sec:conclusions}.

\section{Background on Theory of Scintillation}
\label{sec:theory}

\subsection{Thin screen theory and stationary phase approximation}
\label{sec:scint}
The theory of scattering in thin screens is outlined in detail in \citet{walker+04} and \citet{cordes+06}, and we summarise some of the pertinent relations here.  

The ``stationary phase approximation'' assumes that the observed signal can be described as a coherent summation over all images of the pulsar (stationary phase points, regions where light can be deflected to the observer).  Each image has a geometric time delay $\tau_{i}$ and a fringe rate (or Doppler rate) $f_{D,i}$, with a magnification $\mu_i$ and intrinsic phase $\phi_i$.  In this approximation, the contribution of all of the images is
\begin{equation}
g_{E}(\tau, f_{D}) = \sum_{i} \sqrt{\mu_{i}} e^{-i \phi_{i}}\, \delta(f_{D} - f_{D,i}) \delta(\tau - \tau_{i}).
\label{eq:field}
\end{equation}
What we observe is the intensity as a function of time and frequency $I(\nu, t) = |E(\nu, t)|^{2}$, called the dynamic spectrum, formed using sufficiently fine channels to fully resolve the scintillation in frequency\footnote{equivalently, one must Fourier transform over a long enough timespan of E(t) fully encompassing g(t) (by a factor of 2, due to the Nyquist Theorem) -- the longest timescales correspond to the finest frequencies} and each time bin averaged over many pulse rotations.
The 2D power spectrum of $I(\nu, t)$ is referred to as the secondary spectrum, which expresses the intensity in terms of its conjugate variables $f_{D}$ and $\tau$, and contains the contribution of interference between all \textit{pairs} of images
\begin{equation}
|\tilde{I}(\tau, f_{D})|^{2} \approx \sum_{i,j} \mu_{i} \mu_{j}\, \delta(f_{D} - f_{D,ij}) \delta(f_{D} + f_{D,ij}) \delta(\tau - \tau_{ij}) \delta(\tau + \tau_{ij}),
\label{eq:secspec}
\end{equation}
where $f_{D, ij}$ and $\tau_{ij}$ are the differences between two interfering images, 
\begin{align}
f_{D, ij} &= \frac{ (\bm{\theta_{i}} - \bm{\theta_{j}} ) \cdot \bm{v}_{\rm eff} }{\lambda}, \\
\tau_{ij} &= \frac{d_{\rm eff} (\theta_{i}^{2} - \theta_{j}^{2})}{2 c}.
\end{align}
We note that, since the dynamic spectrum is a real function, the secondary spectrum is point-symmetric.

The effective distance $d_\mathrm{eff}$ and effective velocity
$\bm{v}_\mathrm{eff}$ depend on the fractional distance of the screen from the pulsar $s$ as
\begin{align}
d_{\rm eff} &= (1/s - 1) d_{\rm psr},  \\
\bm{v}_{\rm eff} &= (1/s - 1) \bm{v}_{\rm psr} + \bm{v}_{\oplus} - \bm{v}_{\rm scr} / s, \\
s &= 1 - d_{\rm scr} / d_{\rm psr}.
\end{align}
Where, $d_\mathrm{psr}$ and $\bm{v}_\mathrm{psr}$,  
are the pulsar's distance and velocity, 
$\bm{v}_{\oplus}$ and $\bm{v}_{\rm scr}$ are the velocities of the Earth and scattering screen, respectively (and where we are only considering the 2D velocity on the plane of the sky).

Considering one image as the direct line-of-sight, then $\theta_i$ or $\theta_j=0$, and  $\tau$ and $f_{D}$ are related through their common dependence on $\theta$:
\begin{equation}
\tau = \eta f_{D}^{2}, \quad\mbox{with}\quad \eta = d_{\rm eff} \lambda^{2} / 2c v_{\rm eff, ||}^{2},
\end{equation}
where $\lambda$ is the observing wavelength, and where $v_{\rm eff, ||}$ is the effective velocity along the position vector $\hat{\theta}$ to the image.   For a 1D distribution of images, we denote the angle of the screen's axis with $\psi$.
Many images along a line interfering with the direct line of sight 
then results in a parabolic distribution of power in the secondary spectrum, while the commonly seen ``inverted arclets'' (eg. \citealt{stinebring+01}) arise from the interference between subimages.

The curvature $\eta$ depends on the distance to the screen, the effective velocity, and the angle between the velocity and the screen. Structures in the secondary spectrum move along the main parabola from left to right (negative to positive $f_{D}$) due to the effective velocity as
\begin{equation}
    \frac{d f_{D}}{dt} = \frac{1}{2 \eta \nu} \left(1 - \nu f_{D} \frac{d \eta}{dt} \right).
\end{equation}
The motion of points in the secondary spectrum is uniquely defined by the curvature of the parabolic arc and its time-derivative -- in other words, clumps of power in the secondary spectrum must move, and the resulting bulk scattering time is necessarily variable.  Variable motion from the Earth's or the pulsar's orbit will contribute to $\nu f_{D} \frac{d \eta}{dt}$.

\subsection{The interstellar response}

In this section we summarise and expand upon the method of \citet{hemberger+08}, to use the secondary spectrum to estimate the total time delays from multipath propagation. 

The electric field that we observe is the intrinsic signal of the pulsar convolved with the impulse response function of the ISM, 
\begin{equation}
    E(t) = (E_{\rm int} * g_E)(t),
\end{equation}
where $E_{\rm int}$ is the intrinsic signal of the pulsar, and $g_E(t)$ is the interstellar impulse response function of the field.

We measure the time-averaged intensity, not the direct electric field.  The quantity of interest is then the time shift of the intensity $\langle \tau \rangle_{I(t)}$, where
\begin{equation}
    I(t) = \mean{|E(t)|^{2}} = \mean{|(E_{\rm int} * g_E)(t)|^{2}}, 
\end{equation}
where $\mean{}$ denotes the average over many pulses.
First we must find a suitable way to describe the effect of response function of the field $g_E(t)$ on the intensity.  Under the assumption that the intrinsic field is temporally incoherent, then
\begin{align}
  \mean{E\sub{int}(t_1) E^{*}\sub{int}(t_2)} &= I\sub{int} (t_1) \delta (t_1-t_2),
  \label{eq:incoh}
\end{align}
and it can be shown that the observed intensity can be written as
\begin{equation}
    I(t) = (I\sub{int} * g_{I})(t)
\end{equation}
where $g_I(t) = |g_E(t)|^2$ can be thought of as the intensity response function.  Equation \ref{eq:incoh} is written unrigorously for infinite bandwidth -- in the real case of a finite bandwidth, the delta function would be replaced by a sinc function with width $\sim 1/BW$.

With the above assumptions, we now have the intensity written in a form resembling \citet{hemberger+08}, and can follow their steps.
The goal of this method is to estimate the time shift from the intensity response function, $\langle \tau \rangle_{g_I(t)} \equiv \tau_s$.  Two properties of convolutions are important for this method, that the centroid and the variance of two convolved functions are additive
\begin{equation}
 \langle \tau \rangle_{f * g} = \langle \tau \rangle_{f} + \langle \tau \rangle_{g}, \quad \textrm{and} \quad \sigma^{2}_{f * g} = \sigma^{2}_{f} + \sigma^{2}_{g}
\end{equation}
For simplicity, we define the center of the pulse to be at $t=0$, and define the pulse width to be $w$.
When $w \gg \tau_{s}$ (as is the case for this paper, as the $\sim \mu$s delays are much smaller than the $\sim$\,ms pulse width), then using the two convolution properties above, we have
\begin{equation}
    \langle \tau \rangle_{I} = \langle \tau \rangle_{I_{\rm int}} + \langle \tau \rangle_{g_I} = \tau_s,
\end{equation}
and 
\begin{equation}
    \sigma^{2}_{I} = \sigma^{2}_{I_{\rm int}} + \sigma^{2}_{g_I} \sim w^{2} + \tau_{s}^{2} \approx w^{2},
\end{equation}
since $\langle \tau \rangle_{I_{\rm int}} = 0$ by definition.  This means that the shape of the pulse is effectively unchanged, yet it still has a bulk time delay from the response.  

\subsection{ Estimating time delays from the Secondary Spectrum }

Now we address how to estimate the time delays of the intensity response function in practice.
As we are only concerned with measuring time delays, in this section we drop the time dependence for simplicity, focusing on the imprint of the impulse response function on the spectrum.

The Fourier transform of the intensity spectrum is
\begin{equation}
    \tilde{I}(\nu) = \tilde{I}_{\rm int}(\nu) \tilde{g_I}(\nu),
\end{equation}
where the convolution between the pulsar's signal and impulse response becomes a direct multiplication.  The intrinsic profile $I_{\rm int}$ is assumed to be stable, and only slowly varying across frequency after averaging over many pulse rotations, so we treat it as a constant.  The secondary spectrum is obtained by Fourier transforming and squaring the spectrum $I(\nu)$, resulting in
\begin{equation}
    |I(\tau)|^{2} = I_\text{int}^{2} (g_I(\tau) * g^{*}_I(-\tau)).
\label{eq:Itau}
\end{equation}
This is the autocorrelation of the intensity impulse response function, 
and we see that this form cannot necessarily recover the total time delay as it only measures differences in $\tau$, not an absolute time.  

To simplify, we return to the stationary phase approximation, as discussed in Section \ref{sec:scint}.  The intensity response function is the square modulus of the field as given in equation \ref{eq:field}, where if we assume that the images lose coherence when integrating over the full observation we have
\begin{equation}
    g_I(\tau) = \sum_{i} \mu_{i} \delta(\tau_{i}).
\end{equation}
We wish to estimate this from the secondary spectrum.  To examine a limiting case, let us assume most of the power is near the undeflected line of sight (defined as $j=0$), then $\tau_{0} = 0$, $\mu_0 \approx 1$, and $\mu_0 \gg \mu_i$. Then, taking only positive $\tau$, and averaging over $f_{D}$, equation \eqref{eq:secspec} becomes
\begin{equation}
|I(\tau)|^{2} \approx \sum_{i} \mu_{0} \mu_{i} \delta(\tau_{i}),
\label{eq:Itau_centralimage}
\end{equation}
In this limit, there will be a visibly strong parabolic arc without inverted arclets. 
The total time delay would then be determined from the expectation value in $\tau$, where the contribution of the bright central image divides out
\begin{align}
    \langle \tau \rangle_{I} &= \frac{\sum_{i} \mu_{0} \mu_{i} \tau_{i}}{\sum_{i} \mu_{0} \mu_{i}} \\ &= \frac{\sum_{i} \mu_{i} \tau_{i}}{\sum_{i} \mu_{i} } \\ &= \langle \tau \rangle_{g_I}
    \label{eq:Igequiv}
\end{align}
The contribution of the phases can be neglected if every pixel in the secondary spectrum contains only one pair of interfering images -- while not necessarily the case, this is aided by the time axis of the dynamic spectrum and many channels, which separates the power in the secondary spectrum in $f_{D}$ as well as $\tau$.

We see that in the limit of a strong central image, we can recover the total time delays from the secondary spectrum.
More generically, how well the time delays can be computed from the secondary spectrum is dependent on the unknown distribution of images (or the functional form of $g_I(\tau)$).  In the case of strong scattering, there is no reason to expect a single undeflected line of sight image, but rather there may be many bright, scattered images at small angular separation.  In this case, the time delay computed from the above formula will be overestimated, due to the cross-terms of bright central images interfering.  This will bias the result high by a factor of $\sim 2m / (m+1)$, where $m$ is the number of bright images, leading to a difference as large as a factor of 2.  In the case of a discretised secondary spectrum, this will only begin to matter if the image separations are larger than one pixel in $\tau$, otherwise it will approximate the case of one bright central image.  
Additionally, we are still limited by the fact that the secondary spectrum measures differences in time delays, rather than absolute time delays; if there is a time-shift applied to all images, it would not be captured by our estimate.

With the above caveats mentioned, we use equations \ref{eq:Itau_centralimage} and \ref{eq:Igequiv} as our basis to measure time delays throughout the paper.  These include the assumption of a strong central image, which we believe gives a reasonable estimate for our purposes.  We describe how to compute time delays in practice from our data in Section \ref{sec:delays}, after detailing our data reduction and secondary spectra creation.

\section{Observations}
\label{sec:data}

\subsection{LEAP}

The Large European Array for Pulsars (LEAP) is a phased array of
five large radio telescopes in Europe; the Effelsberg telescope,
the Lovell telscope at Jodrell Bank Observatory, the Westerbork
Synthesis Radio Telescope, the Nan\c cay Radio Telescope and the
Sardinia Radio Telescope \citep{stappers+11}. The coherent addition of radio signals
from all these telescopes results in an effective 195\,m diameter
dish. The overview of LEAP is given in \citet{bassa+16}.
Observations have been made  monthly since 2012, with whichever subset of these telescopes was available.  

The voltage data from each site are shipped or transferred to Jodrell Bank Observatory to be correlated and coherently added on a designated CPU cluster, using a specifically designed software correlator (details in \citealt{smits+17}).  Correlation involves a polarisation calibration based on an observation of PSR~J1022+1001 or PSR~B1933+16 from the same epoch, correlation on a calibrator to find an initial phasing solution, then self-calibration on the pulsar to determine the time delays and fringe drift rates for each telescope
throughout the observation, using Effelsberg as a time and position reference.  The coherently added voltages are stored on tape, allowing us to re-reduce the data with arbitrary time or frequency resolution.  
The high sensitivity, and the flexibility offered by storing the baseband data has enabled LEAP to do single pulse studies of MSPs \citep{liu+16, mckee+19}; for these same reasons, it is an ideal telescope for the scintillation work presented in this paper.  Typical observing lengths are $30-60$\,minutes, with bandwidths of $80-128$\,MHz (comprised of 16\,MHz subbands), depending on the subset of telescopes used for a given observation.
As we will show in Section \ref{sec:screen}, the angular extent of the scattering screen is unresolved by LEAP, so we can safely treat it as a single-dish instrument for our purposes.

\subsection{Effelsberg 100--m Telescope}
\label{sec:effelsberg}

From March to June 2020, we had a roughly bi-weekly monitoring campaign using the Effelsberg telescope.  Baseband data were recorded as 8-bit ``dada''\footnote{\url{http://psrdada.sourceforge.net/}} files using the PSRIX backend (described in \citealt{lazarus+16}), using the central feed of the 7-beam receiver (``P217mm'').  The data were recorded in 25\,MHz subbands, with a usable bandwidth of 1250-1450\,MHz, and typical observation lengths of $90\,$minutes.  While Effelsberg alone is less sensitive than LEAP, this is compensated through the larger exposure times and bandwidth.  

\section{Methods}
\label{sec:methods}

\subsection{Creating dynamic and secondary spectra}

We created folded archives from the baseband data using \texttt{dspsr} \citep{vanStraten+11}, coherently de-dispersing and folding with 10\,s bins, 128 phase bins, and sufficient channels to fully resolve scintillation - 62.5\,kHz, and 50.0\,kHz channels for LEAP and Effelsberg respectively. The subbands were combined in frequency using the \texttt{psrchive} tool \texttt{psradd} \citep{hotan+04} to form one combined archive per observation. The following processing steps for data from either telescope are identical unless expressly stated otherwise.

After summing polarisations, each folded archive contains a data cube $I(t, \nu, \textrm{phase})$.  
We use a fixed off-pulse region relative to the pulse, a contiguous $50\%$ section with no apparent pulsed emission (in Figure \ref{fig:template}, phase 0.5--1.0)
We divide by the time average of the off-pulse region across the full observation to approximately remove the bandpass, and in each time and frequency bin, we subtract the mean of the off-pulse region to remove variable background flux. Sub-integrations with an off-pulse standard deviation $>5\times$ the mean rms value were masked, as were any time bins or frequency channels with $>30\%$ of flagged sub-integrations.

The LEAP dada files are saved separately in each sub-band, in individual 10\,s files; a small number of these files were missing, and were filled with zeros, and included in our mask.  To reduce artefacts caused by Fourier transforming over a window function, masked pixels were iteratively in-painted using the mean of the nearest pixels.  While more sophisticated methods of inpainting exist, this is sufficient for our analysis, as typically no more than $5\%$ of data are flagged.

\begin{figure*} 
\center{}
\includegraphics[width=1.0\textwidth,trim=0cm 0.0cm 0cm 0cm, clip=true]{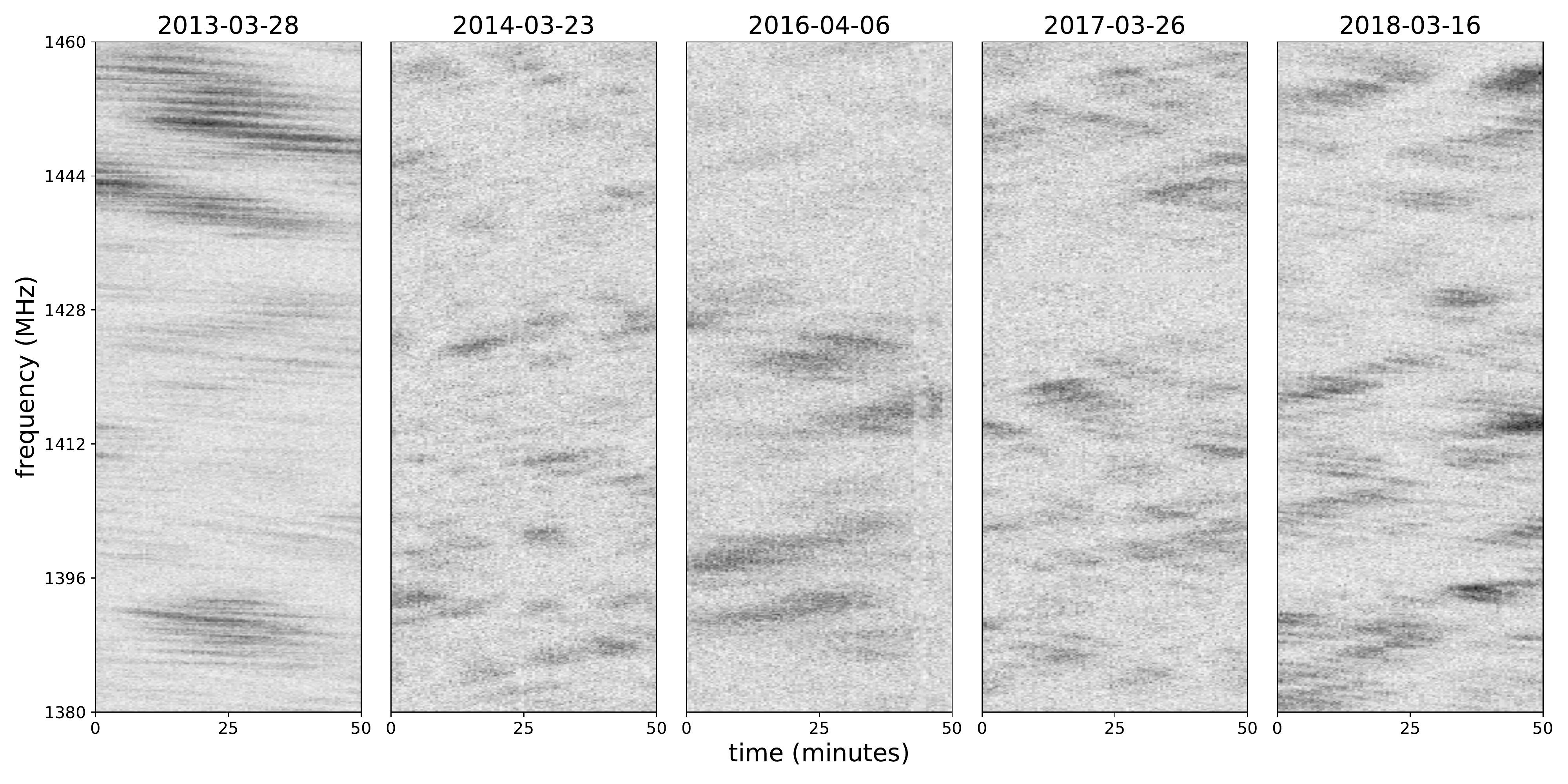}\\
\includegraphics[width=0.992\textwidth,trim=-0.55cm 0cm 0cm 0.0cm, clip=true]{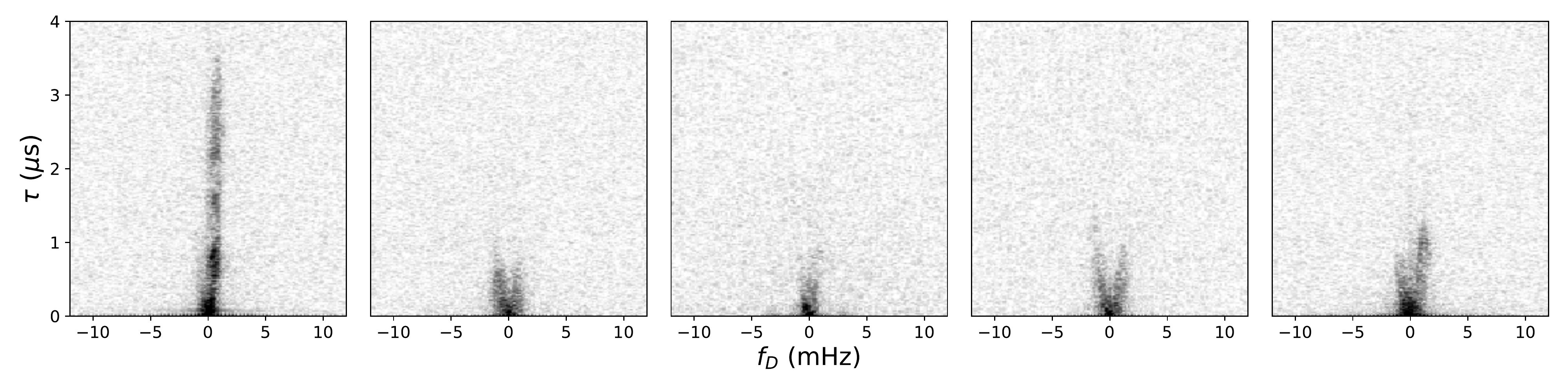}
\vspace{-3mm}
\caption[Dynamic Spectra of LEAP pulsars]{{\it Top:} Dynamic Spectra of 5 observations around the same time of year, to have comparable contributions from the Earth's velocity. The colourbar extends from $2\sigma$ below the mean to $5\sigma$ above. {\it Bottom:} Corresponding secondary spectra, with a logarithmic colourbar extending three orders of magnitude.  Clear arcs with noticeable localised clumps of power are seen, these correspond to prominent diagonal features in the above dynamic spectra.  The observation from 2013 is anomalous, showing extremely fine stripes in the dynamic spectrum, corresponding to power at large time delays.
\label{fig:dynspec}}
\end{figure*}

A time and frequency averaged profile was created, and zeroed everywhere the S/N was below 5$\,\sigma$.  This profile was used to weight each phase bin, before summing over pulse phase to create the dynamic spectrum I(t, $\nu$).  
Over a narrow band, it is sufficient to simply use a 2D FFT, which we used for this analysis (over a wider band, the $\nu^{-2}$ scaling of $\eta$ causes arcs to smear in the secondary spectrum, summarised in \citealt{gwinn+19}).
Before taking a FFT, we padded the edges by a factor of two with the mean value of the dynamic spectra, to mitigate artefacts caused by edge effects. 
A few representative LEAP dynamic and secondary spectra, at the same time of several years, are shown in Figure \ref{fig:dynspec}. 

\subsection{Measuring arc curvatures}
\label{sec:curvature}

The main power in the secondary spectrum of \psr follows a parabolic arc, suggesting scattering dominated by a highly anisotropic, thin-screen.  As described in Section \ref{sec:screen}, the arc curvature encodes $v_{\rm eff}$ and $d_{\rm eff}$; we wish to measure the arc curvature for each observation, to probe the changing velocities of the system, and to localise power for measuring time delays.
One method often used to measure parabolic curvatures is the Hough transform, finding the peak of the power summed over different possible parabolic curvatures \citep{bhat+16}.  While this technique works very well for thin parabolic arcs, it leads to a broad maximum (and correspondingly large uncertainties) for broad parabolae, as seen in our data. We determine the curvature in a different way;  in $\tau$ steps of $0.125\,\mu$s in the secondary spectrum (where $\tau > 0.5\,\mu$\,s, to avoid confusion in the bright centre), we find the peak value of $I(f_{D})$ for both positive and negative $f_{D}$. We keep only points where the peak is $>4\times$ the rms of the background, estimated from the region of $I(|f_{D}| > 10\,\mu\text{s})$. This set of points in $f_{D}$ and $\tau$ is fit with a parabola, using an orthogonal minimisation routine, to find the best-fit curvature and error.

\subsection{Integrating the secondary spectrum}
\label{sec:delays}

For purposes of measuring time delays, the $x$-axis $f_{D}$ is not important, except to localise the scattered power in this parameter space.  We isolate the power in a $1\,$mHz region surrounding the main arc, as defined by our measured arc curvatures.
We subtract the averaged background far from the main arc, assuming the noise is well described as a function of time-delay.  We measure the total time delay through the expectation value of $\tau$, computed as
\begin{equation}
\langle \tau \rangle = \frac{ \int_{0}^{T} \tau |I(\tau)|^{2} d\tau}{\int_{0}^{T} |I(\tau)|^{2} d\tau},
\label{eq:tauint}
\end{equation}
where $T = 8 \mu$s, defined by our choice of channelisation.

Artefacts in the dynamic spectrum, such as RFI, phasing imperfections, and the window function lead to correlated features in the secondary spectrum.  As such, the noise properties are not always well behaved, and direct error propagation underestimates the error on $\langle \tau \rangle$.  We estimate our errors directly from the cumulative function in equation \eqref{eq:tauint}; at high enough $T$ the integral plateaus, with residual variations caused by the effect of integrating noise in the secondary spectrum. We take the mean and standard deviation of equation \eqref{eq:tauint} between $T=4-8\,\mu$s as our measurement and error of $\langle \tau \rangle$ respectively.

\subsection{"Timing" a convolved template}
\begin{figure} 
\includegraphics[width=1.0\columnwidth,trim=0cm 1.0cm 0.0cm 1.0cm, clip=true]{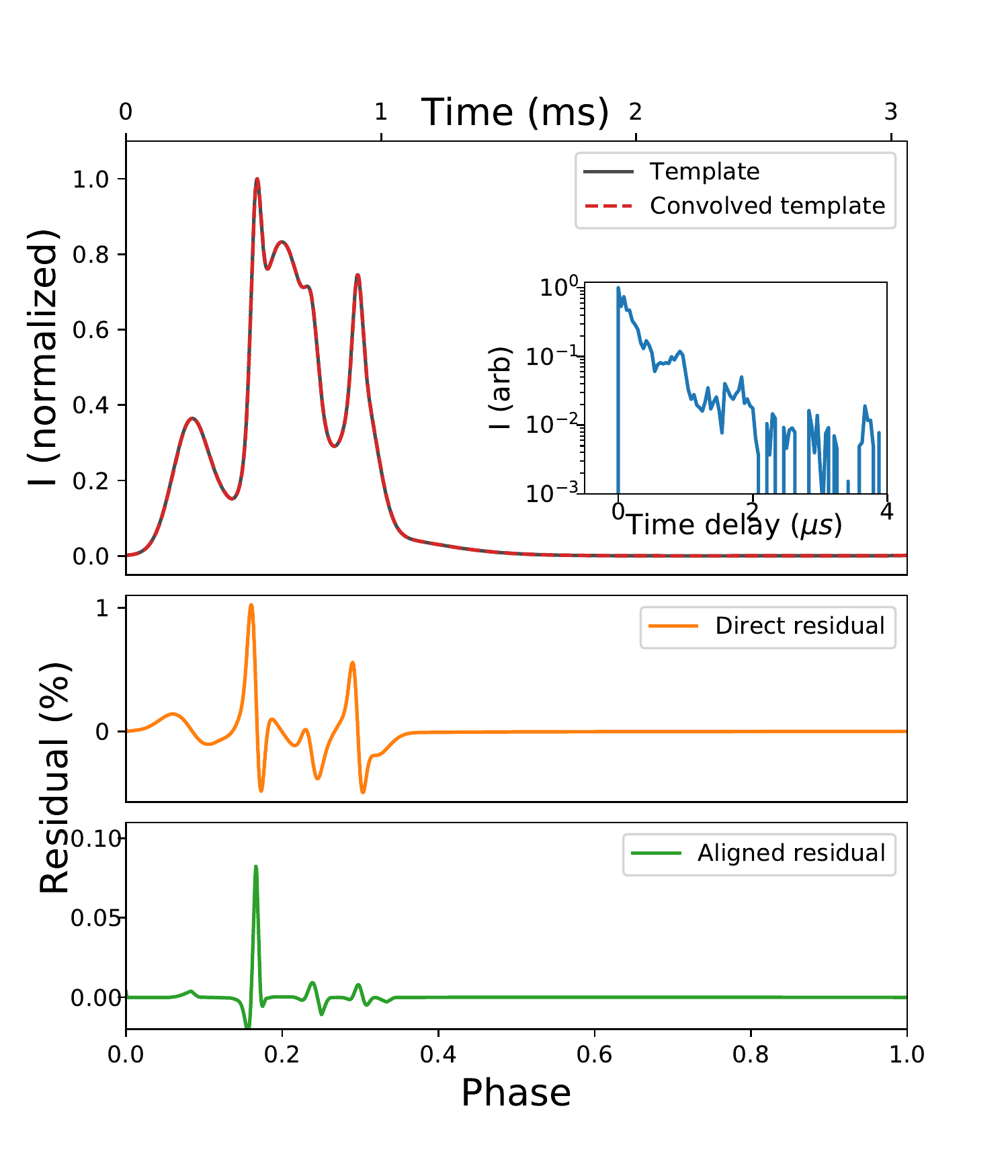}
\caption[Template convolved with scattering tail]{The effects of scattering on a pulse profile. {\it Top:} Analytic profile, before and after convolving with the scattering tail measured from scintillation.  The inset panel shows $\tilde{I}(\tau)$ measured on 2017-02-17 in the way described in Section \ref{sec:delays}.  The resulting convolved profile looks identical by eye to the original, as the time delays are largely sub-bin.  {\it Middle:} The residual of the original and convolved profile, {\it Bottom:} 
the residual between the template and shifted profile, after shifting them into closest alignment.
The middle, uncorrected residuals are only as large as $1\%$, while after aligning the residuals are below the $0.1\%$ level.  This reinforces the fact that the main contribution of scattering is a bulk time shift, rather than a noticeable change in profile shape.
\label{fig:template}}
\end{figure}

To illustrate the effects of scattering on a profile, 
we can directly convolve our measure of the amplitude of $g_I(\tau)$ into a template profile and measure the time offset using the standard Fourier template-matching algorithm outlined in the appendix of \citet{taylor92}.  We create an analytic template using the standard \texttt{psrchive} tool \texttt{paas} \citep{hotan+04}, fitting the profile with a series of von Mises functions, and interpolate the solution to have the equivalent 31.25\,ns bins of our measured $|I(\tau)|^{2}$.  We convolve the two, and measure the relative time delay between the convolved template against the original one.  Figure \ref{fig:template} shows this convolution applied to one of our observations.  The measured time delay in this way agrees perfectly with the method in the previous section; timing recovers the shift correctly, even when the effects are not visibly noticeable.

We note again that this is not precisely the intensity impulse response, but rather its autocorrelation, but it is close enough in amplitude to demonstrate that the convolved template is visually identical (with residuals at the $0.1\%$ level after aligning the template), yet is measurably delayed. In addition, $|I(\tau)|^{2}$ is noticeably clumpy and poorly described by an exponential, even after being effectively smoothed by the autocorrelation.

\begin{figure} 
\center{}
\includegraphics[width=1.0\columnwidth,trim=0cm 0cm 0cm 1.5cm, clip=true]{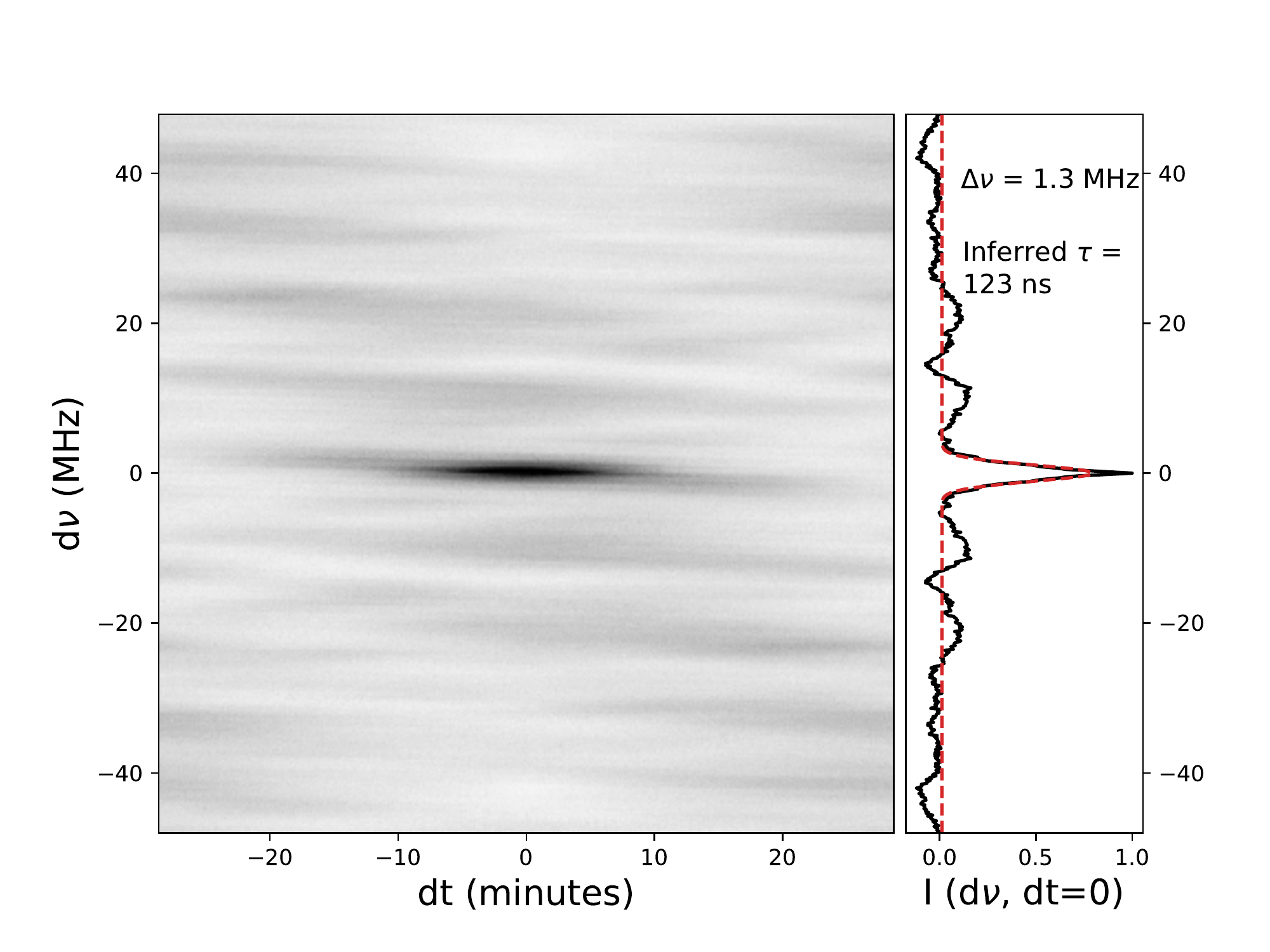} \\ 
\includegraphics[width=1.0\columnwidth,trim=0cm 0cm 0cm 1.5cm, clip=true]{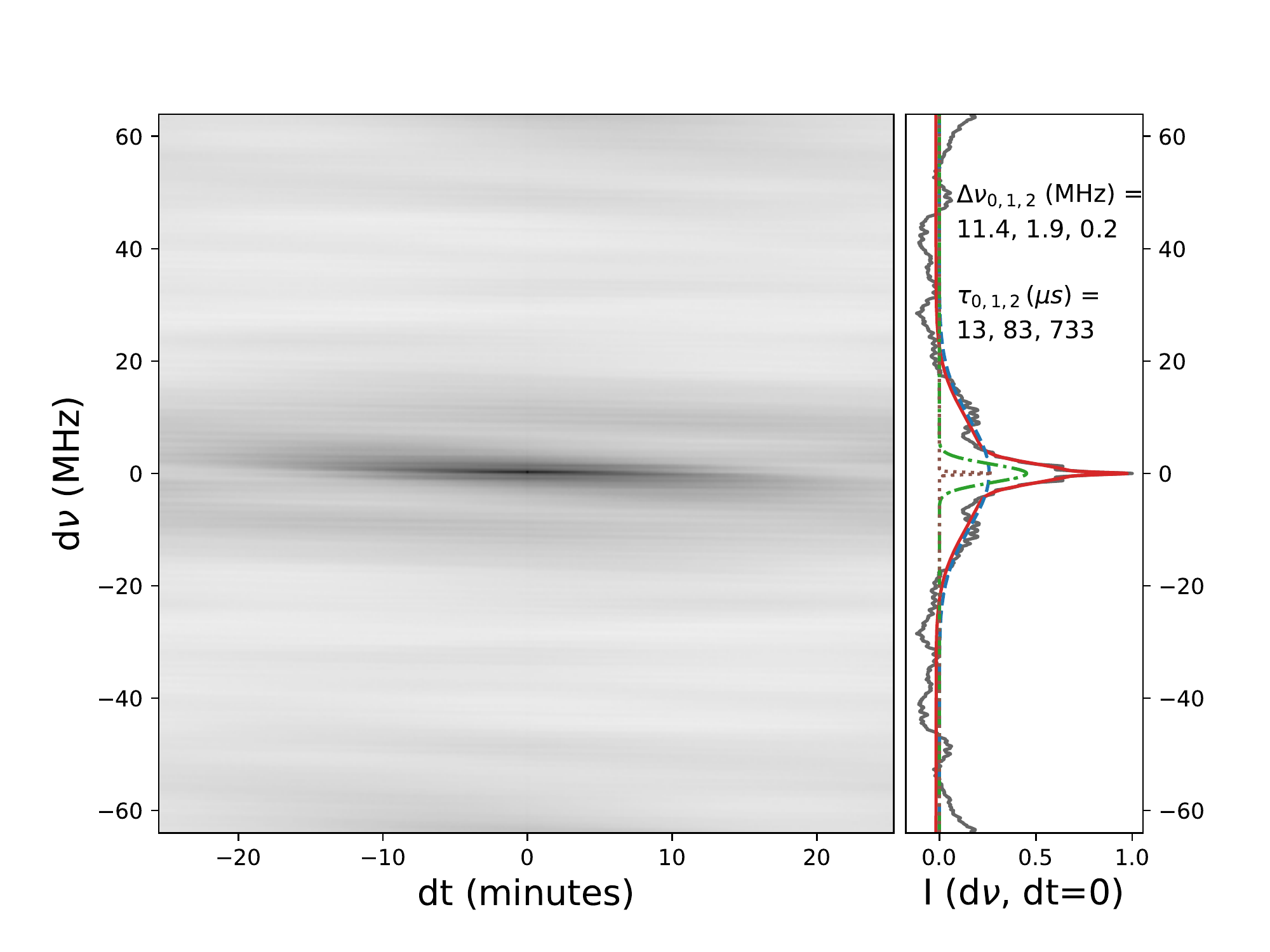}
\vspace{-5mm}
\caption[Cross correlation]{Two examples of 2D ACFs.  The cut through dt=0 is fit with a Gaussian to measure the scintillation bandwidth, to infer the bulk scattering time. {\it Top:} the same data as in Figure \ref{fig:template}, well fit by a 1D Gaussian.  {\it Bottom:} ACF of the leftmost panel of Figure \ref{fig:dynspec}, showing 3 distinct frequency scales. 
\label{fig:acf}}
\end{figure}

\subsection{Inferred time delay from the frequency ACF} 
\label{sec:acf}
A standard way to infer the time delays from scattering is to construct the auto-correlation functions $R(\Delta \nu) = (I * I) (\Delta \nu)$.  Fitting the width (specifically, the HWHM) of the ACF in frequency gives the scintillation bandwidth $\nu_{\rm scint}$, which is inversely proportional to the bulk scattering delay as $\langle\tau \rangle = C / 2\pi\nu_{\rm scint}$ (C is commonly assumed to be 1, and depends on the assumptions of the scattering distribution).  This method is often used when arcs cannot be resolved nicely, as $\Delta \nu$ can typically be measured simply and robustly.  However, if $g_E(t)$ is not smooth, then the ACF will be poorly described as a single Gaussian.  Two such examples of an ACF, one well described, and one poorly described by a 1D Gaussian fit are shown in Figure \ref{fig:acf}, along with their derived $\nu_{\rm scint}$ and $\langle \tau \rangle$.  In the second case, a single Gaussian would preferentially fit the broad component, and result in an inferred time delay which is low, while power at large time delays results in the narrow peak smaller than 1\,MHz. 

Our error includes both the measurement error of the fit, in addition to the ``finite scintle error'', which is a counting error of $\sqrt{N_{\rm scintles}}$, estimated in the same manner as \citet{levin+16} as 
\begin{equation}
    \delta \langle \tau \rangle / \langle \tau \rangle \approx [( 1 + \eta_{t} T_{\rm obs} / t_{\rm scint}) (1 + \eta_{\nu} BW / \nu_{\rm scint})]^{-1/2}.
\end{equation}
The values of $\eta$ are the filling fraction of scintles, assumed here to be $0.2$.

\section{Results}
\label{sec:results}

\begin{figure*}
\center{}
\includegraphics[width=1.0\textwidth,trim=0cm 0cm 0cm 0cm, clip=true]{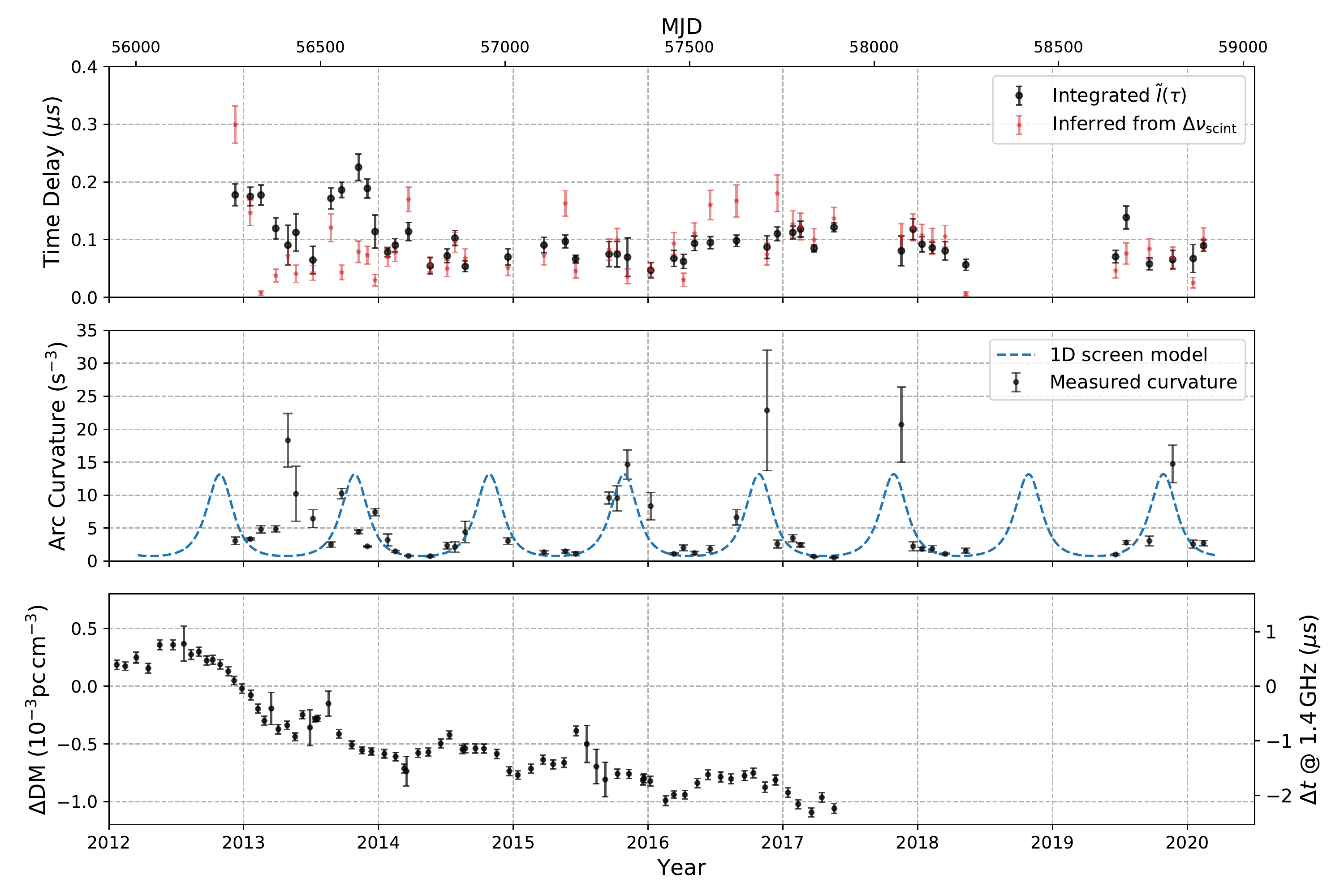}
\vspace{-7mm}
\caption[Cross correlation]{{\it Top:} Variations in the bulk scattering time from every LEAP observation.  Black points are estimated through integrating $\tilde{I}(\tau)$, as described in Section \ref{sec:delays}. The red points are delays inferred from the measurements of $\nu_{\rm scint}$ from the ACF, described in Section \ref{sec:acf}. 
{\it Middle:} The measured arc curvatures for each observation, with a best-fit model of a 1D scattering screen from 2014 onwards overplotted.  {\it Bottom:}  Measured DM values from NANOGrav 12.5 year data release \protect{\citep{nanograv20a}}.
\label{fig:scattering}}
\end{figure*}

\subsection{Evolution of scattering time from LEAP: month to year timescales}

We present our measurements of the bulk time delay in the top panel of Figure \ref{fig:scattering}.  We find significant persistent scattering at the $\sim80\,$ns level, and a few cases of strong scattering variability on several month to year timescales.  The timescales are set by the time it takes power to move through the secondary spectrum, as described in Section \ref{sec:theory}.  The most striking feature is the strong excess scattering in 2013, where the bulk scattering is variable and extends above $200\,$ns.  This is not captured very well by the ACF method, as the Gaussian fit latches onto the broad-scale scintillation rather than the narrow peak caused by the large time delays, as described in Section \ref{sec:acf}.  As hinted at in Figure \ref{fig:acf}, this could potentially be remedied by using a multi-component model to the ACF, as in principle the ACF contains the equivalent information as the secondary spectrum, only differing by a Fourier transform.

In the bottom panel of Figure \ref{fig:scattering}, we plot the DM values of \psr from NANOGrav's 12.5 year release \citep{nanograv20a}. The DM is steeply decreasing prior to 2013, and there is clear annual variation; this was studied in detail in \citet{jones+17}, with data spanning from 2006 until near the end of 2013.  The authors fit the time variations of DM with a 1-year period sinusoid and a linear trend, to capture the contribution of the pulsar's observed trajectory through the ISM from the pulsar's and Earth's velocity, respectively.  The residuals of the DM show a borderline significant dip of $2-3 \times 10^{-3}$\dmu at MJDs 56300--56400, the beginning of 2013.

Although the ecliptic latitude of the pulsar is quite large ($-25.4^{\circ}$), part of the annual variation that appears in the DM time series might be explained with the contribution of the Solar wind. The time of closest approach for \psr is in mid-June each year, and by modelling the distribution of electrons in the Solar wind as spherical \citep{edwards+06} with an electron density of 7.9 cm$^{-3}$ at the Earth orbit \citep{madison+19}, we expect a DM displacement in the order of $2\times10^{-4}$\dmu at the closest approach. By allowing the amplitude of the Solar wind approximation to vary year by year, we find a model that accounts for both the ISM and the Solar wind is preferred in 7 years across the dataset, while an ISM-only model favoured elsewhere. The years in which the complete model is preferred show a general compatibility with an amplitude of 7.9 cm$^{-3}$, and after the subtraction of the time-dependent approximation of the Solar wind the most significant remaining feature is the steep gradient of the DM leading into 2013.

Similar events of increased scattering have been seen in PSR J1017-7156 and PSR J1603-7202 by the Parkes Pulsar Timing Array (PPTA), in which the scintillation bandwidth and timescale decrease suddenly associated with a jump in DM of several $10^{-3}$\dmu, interpreted as an extreme scattering event \citep{coles+15}.  While we do not see an increase in DM of this order, the increased scattering we observe in \psr may be of similar origin.

\subsection{Location and nature of the scattering screen}
\label{sec:screen}

As mentioned in \ref{sec:delays}, we fit the parabolic curvature of each observation to determine the masks for estimating the time delays. The arc curvatures contain the effective velocity, and vary throughout the year from the Earth's motion. The existence of parabolic arcs suggests highly anisotropic scattering; for a one-dimensional screen, the arc curvature then depends only on the effective velocity parallel to the screen.  By measuring the change in arc curvature over the year, one can measure the distance and orientation of the scattering screen.  

We perform only a simple analysis here, currently ignoring the contribution from the pulsar's orbital motion.  We fit the observed curvature values beyond 2013 with a 1-dimensional screen, using measured values of the pulsar's distance of 780$\pm80$\,pc and proper motion of $\mu_{ra}=1.822 \pm 0.008$\,mas/yr, $\mu_{dec}=-10.355 \pm 0.017$\,mas/yr from \citet{desvignes+16}.  The three free parameters are the fractional screen distance $s$, the orientation of the screen $\psi$, and $v_{\rm ism,||}$, the velocity of the scattering screen parallel to its axis of anisotropy.  A 1D screen fits the data well, shown in the middle panel of Figure \ref{fig:scattering}, while the best fit values are in Table \ref{table:fitresults}.
Using the screen distance, and the largest detectable time delays of $\tau \approx 5\,\mu$s, the largest angular extent of the screen is $\theta \approx \sqrt{2c \tau / d_{\rm eff}} \approx 3$\,mas, smaller than the resolution of the longest baselines of LEAP.

\begin{figure*} 
\center{}
\includegraphics[width=1.0\textwidth,trim=0cm 0.cm 0cm 0cm, clip=true]{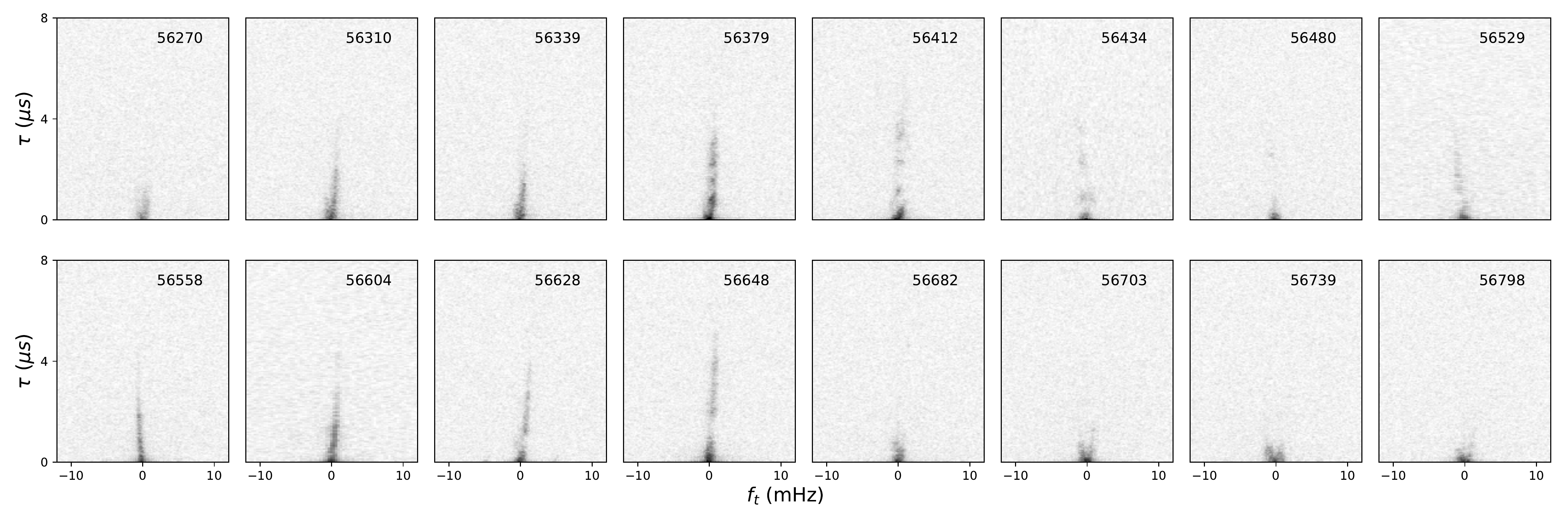}\\
\vspace{-3mm}
\caption[Secondary Spectra of 2013 anomalous scattering]{ Secondary spectra of all observations covering the anomalous scattering in 2013, with a logarithmic colourbar extending four orders of magnitude. The power can be seen to cross the $f_{D}=0$ line, indicating the sign of $\bm{v}_{\rm eff}$ is changing throughout the year, and power can be seen to travel from left to right along the parabola.
\label{fig:CS2013}}
\end{figure*}

\begin{figure} 
\center{}
\includegraphics[width=1.0\columnwidth,trim=0cm 0.0cm 0cm 0cm, clip=true]{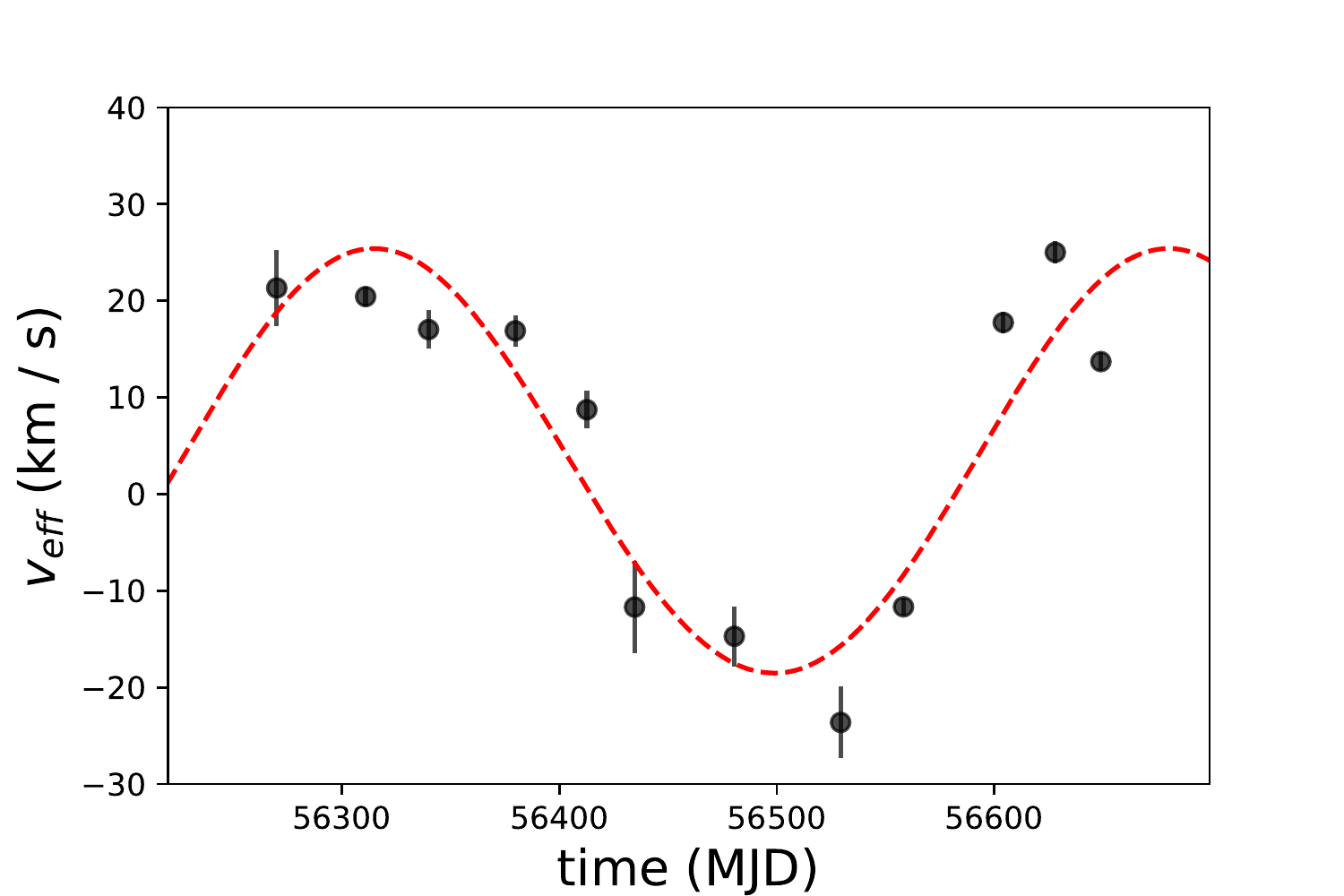}
\caption{ Effective velocity of 2013 scattering event, and best fit 1D model. The sign of $v_{\rm eff}$ was assigned to switch when the dominant power in the secondary spectra was clearly seen to cross the $f_{D} = 0$ axis.
\label{fig:veff2013}}
\end{figure}

During the increased scattering of 2013, the best-fit model poorly matches the data.  Here, we investigate this year separately.  The secondary spectra spanning 2013 are shown sequentially in Figure \ref{fig:CS2013}.  Persistent clumps of power can be tracked throughout the year, and features can be seen to cross the line of $f_{D} = 0$, indicating a changing sign of $\bm{v}_{\rm eff}$.  From this, we can fit directly the signed value of $1/\sqrt{\eta} \propto v_{\rm eff}$ (with the same free parameters $s$, $\psi$, and $v_{\rm ism,||}$), where the locations of velocity zero-crossings are quite constraining.  The measures of $\bm{v}_{\rm eff}$, and best fit model are shown in Figure \ref{fig:veff2013}.

The best fit screen parameters for 2013, and for all data beyond 2013 are tabulated in Table \ref{table:fitresults}.  The distance of the screen is consistent between both fits, with the orientation and parallel screen velocity differing between the two.  This implies that the strong scattering plausibly arises from the same physical region. In addition, the absolute velocity of the screen need not be changing, as the orientation differs and we are only sensitive to the component of the screen velocity parallel to $\psi$; the results of both fits are consistent with a screen velocity of $|v_{\rm scr}| = 15\pm2$\,km/s at an angle of $\phi_{\rm vel} = 15^{\circ} \pm 10^{\circ}$ (East of North).

We note however that the models above are incomplete, as a proper treatment needs to include the binary motion of the pulsar, which we have neglected. The orbit is 1.2 days, and $v_{orb} \sin(i) = 19.9\,$km\,s$^{-1}$.  Each observation is much smaller in duration than the orbit, but is at an effectively random orbital phase. This will add scatter in the velocities, and thus the curvatures. To properly account for the orbital velocity, one would need to jointly fit for $i$ and $\Omega$ - such a fit has been performed successfully on 16 years of arc curvature values of J0437-4715 in \citet{reardon+20}, measuring the inclination with $0.3^{\circ}$ precision and the longitude of ascending node with $0.4^{\circ}$ precision.
Regardless, a 1D screen is a good fit to the data beyond 2013 (where each year the curvature peaks around November and is minimal around May), where the annual variation is the strongest effect. 

\begin{table} 
    \caption{Fit parameters of 1D screen to the arc curvatures, as defined in Section \ref{sec:screen} }
    \centering
    \begin{tabular}{|c|c|c|c|}
    \hline
     & $s$ & $\psi_{\rm scr}$\ (degrees) & $v_{\rm ism,||}$\ (km/s) \\
    \hline
     2014 onwards & $0.62 \pm 0.06$ & $16 \pm 2$ & $-1.2 \pm 2.5$ \\
    \hline
     2013 event & $0.58 \pm 0.10$ & $-36 \pm 9$ & $12.8 \pm 2.8$ \\
    \hline
    \end{tabular}
    \label{table:fitresults}
\end{table}

\subsection{Arclet evolution with Effelsberg: Scattering time on week to month timescales}

\begin{figure*} 
\center{}
\includegraphics[width=1.0\textwidth,trim=0cm 0.0cm 0cm 0cm, clip=true]{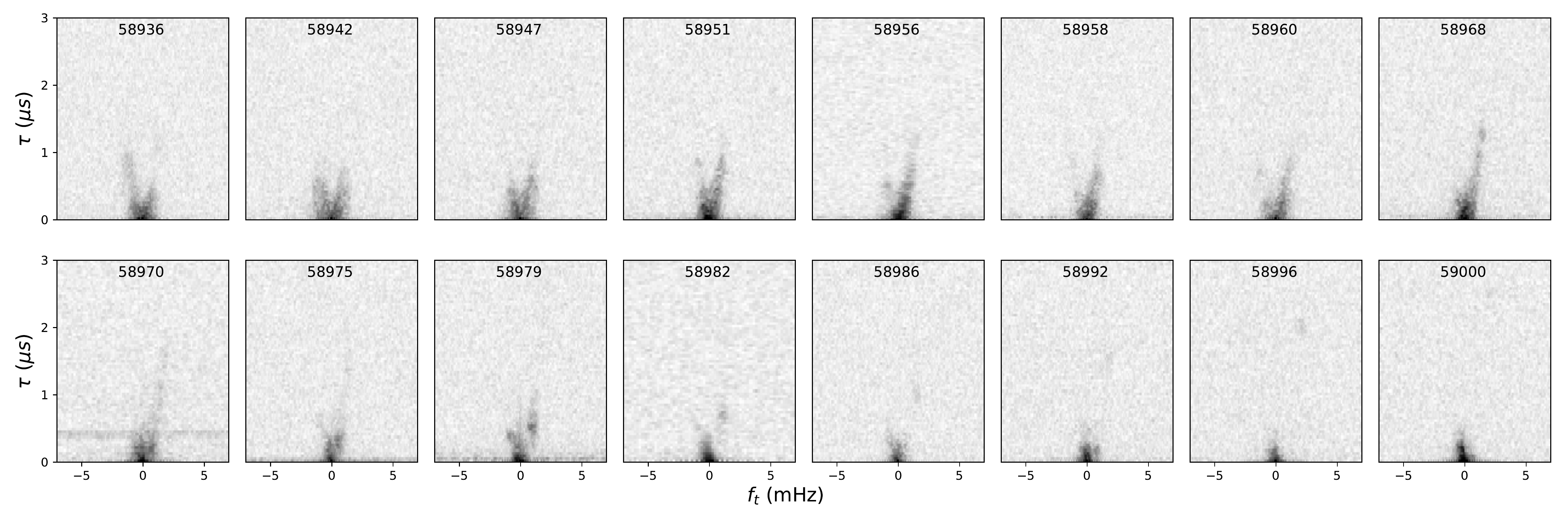}\\
\vspace{-3mm}
\caption[Secondary Spectra of 2020 weekly monitoring]{ Secondary spectra of our roughly bi-weekly monitoring campaign with Effelsberg, with a logarithmic colourbar extending three orders of magnitude.  Power can be seen to travel from left to right along the parabola, most evident by following the power in the last six panels.  The arc curvature does not abruptly change between observations, despite being at random orbital phases, suggesting that the scattering screen resulting in these arcs is not very sensitive to the pulsar's orbit.
\label{fig:CS2020}}
\end{figure*}

\begin{figure} 
\center{}
\includegraphics[width=0.9\columnwidth,trim=0cm 0.3cm 0cm 1cm, clip=true]{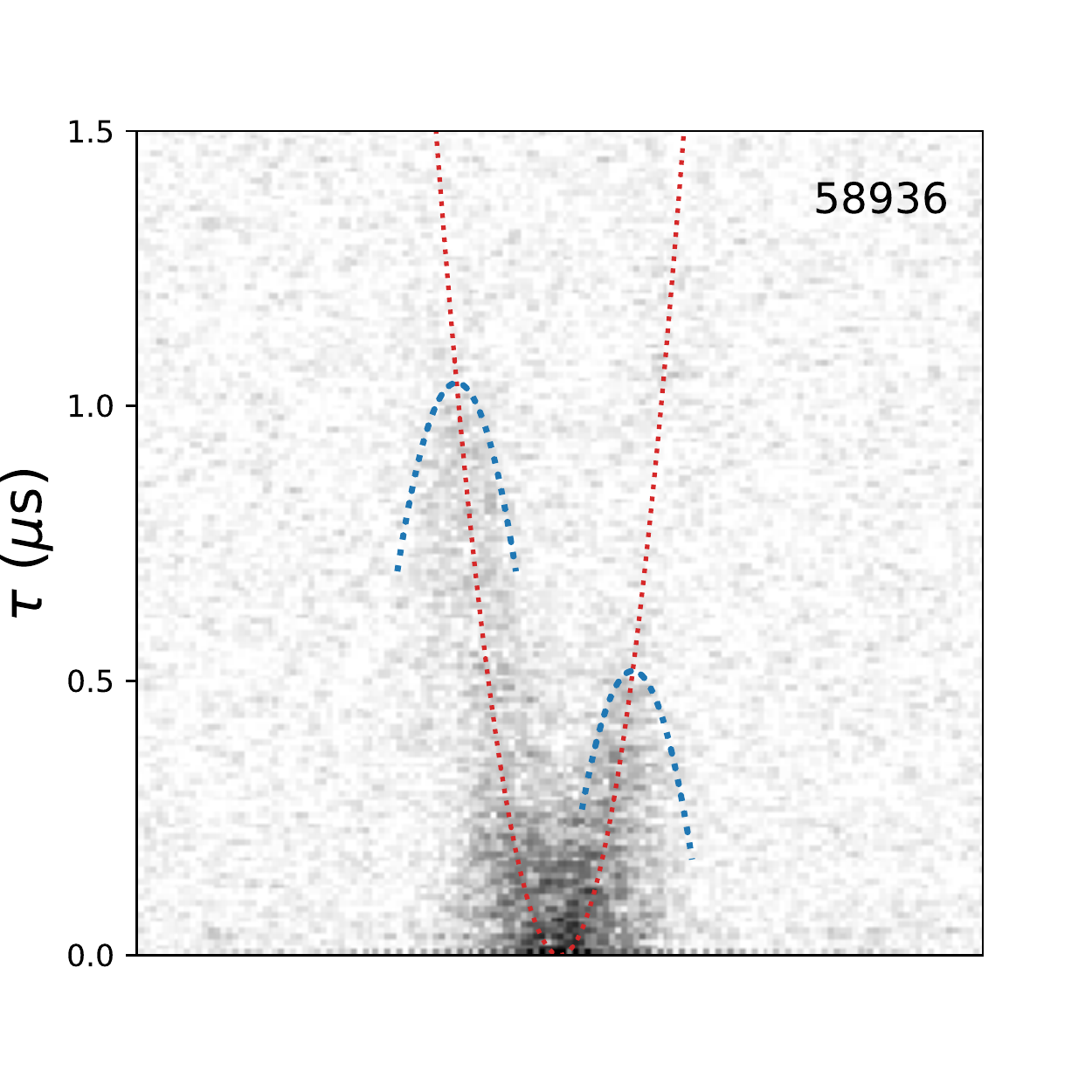}\\
\vspace{-5mm}
\includegraphics[width=0.9\columnwidth,trim=0cm 0.3cm 0cm 1cm, clip=true]{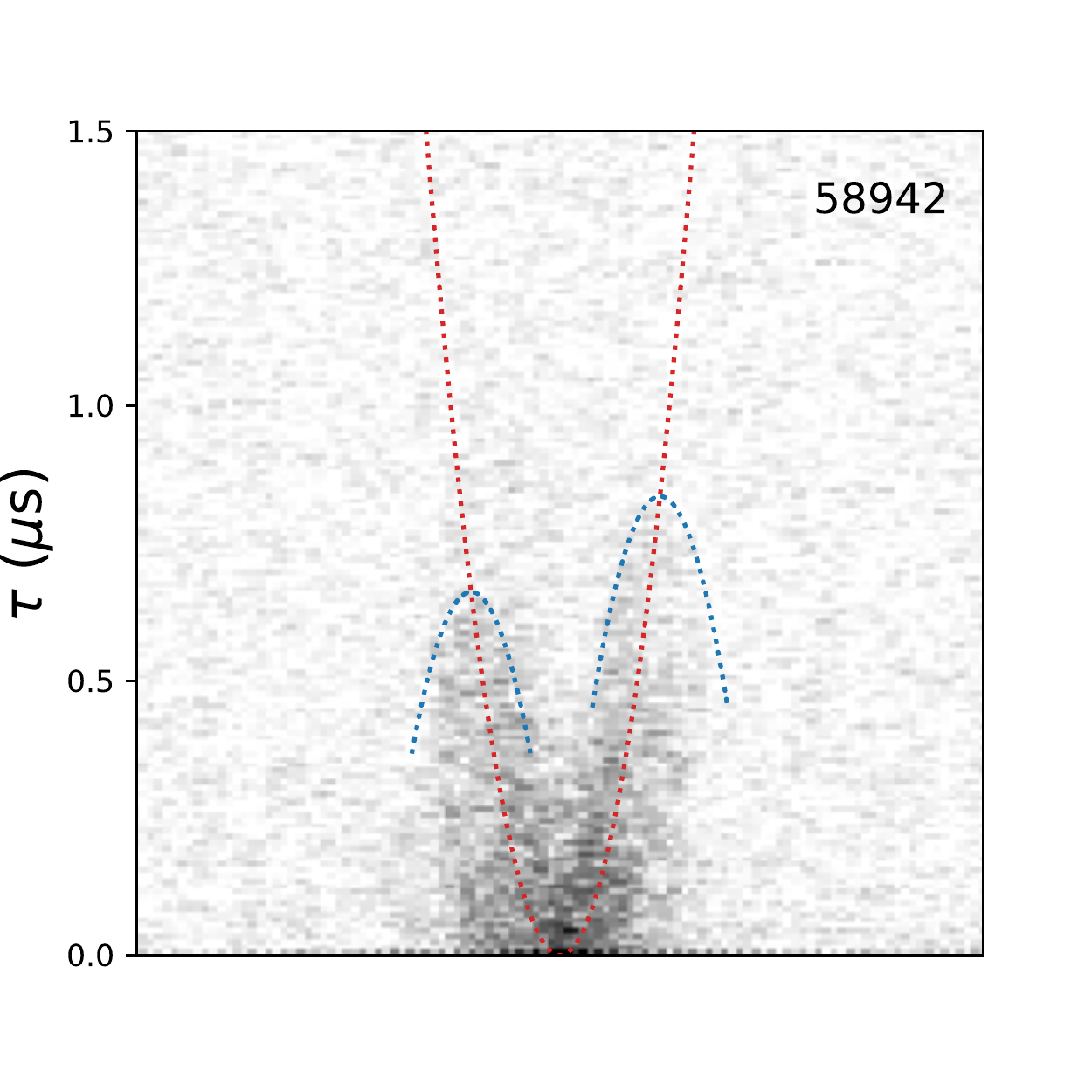}\\
\vspace{-5mm}
\includegraphics[width=0.9\columnwidth,trim=0cm 0.3cm 0cm 1cm, clip=true]{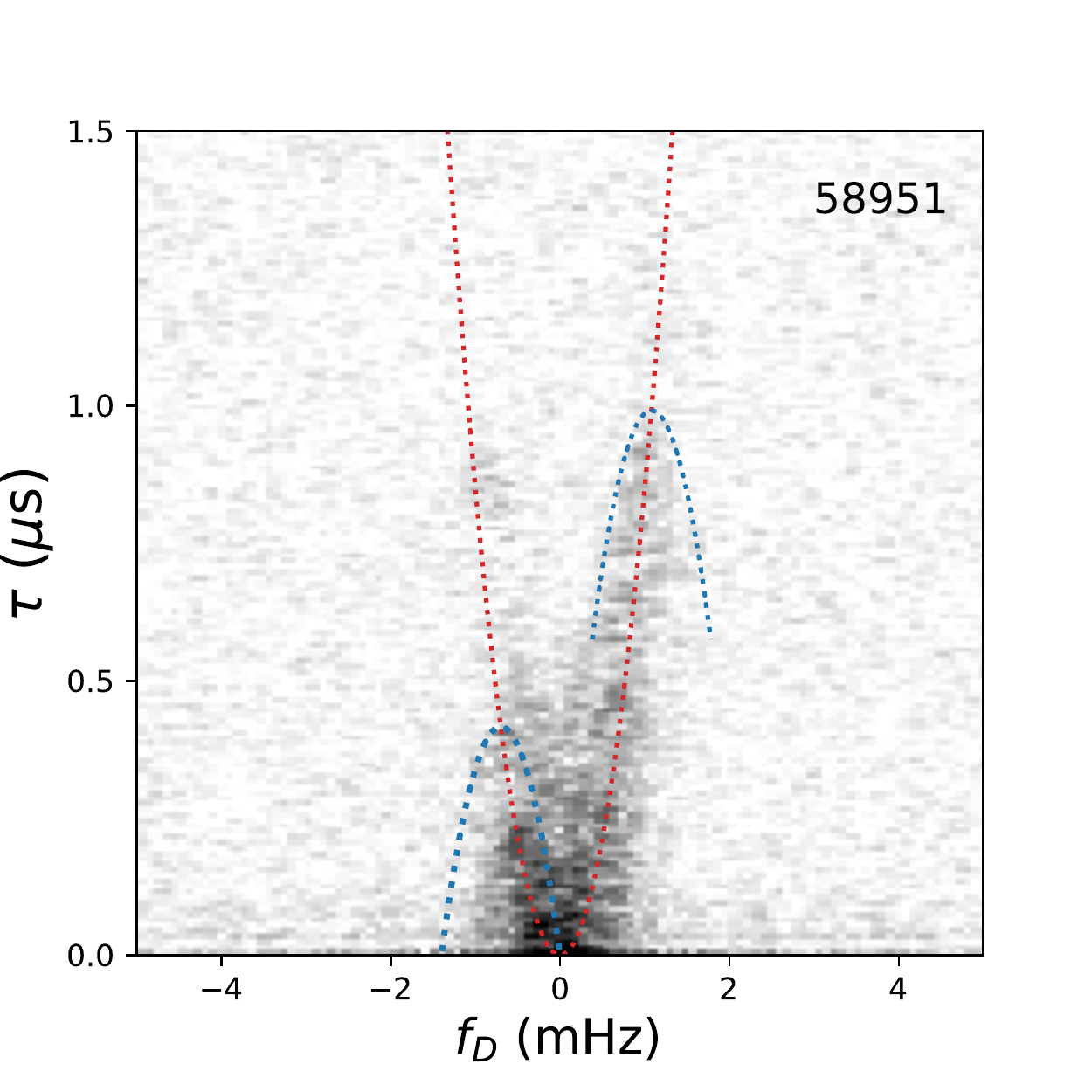}
\vspace{-2mm}
\caption[Secondary Spectra zooms]{ Zoomed in Effelsberg secondary spectra from Figure \ref{fig:CS2020} for three different epochs, panels 1, 2 and 4.  The red dotted line shows our best fit parabolic curvature, while blue dotted inverted parabolae are plotted with the same curvature, with their apex on the main parabola -- these are to guide the eye towards faint structures in the secondary spectra which may be inverted arclets, and to help visualise the general trend of power moving from left to right along the parabola. More sensitive, or longer observations will be required to reveal the possible structure of inverted arclets more clearly.
\label{fig:SecSpecZoom}}
\end{figure}

\begin{figure}
\center{}
\includegraphics[width=1.0\columnwidth,trim=0cm 1.5cm 0cm 1.5cm, clip=true]{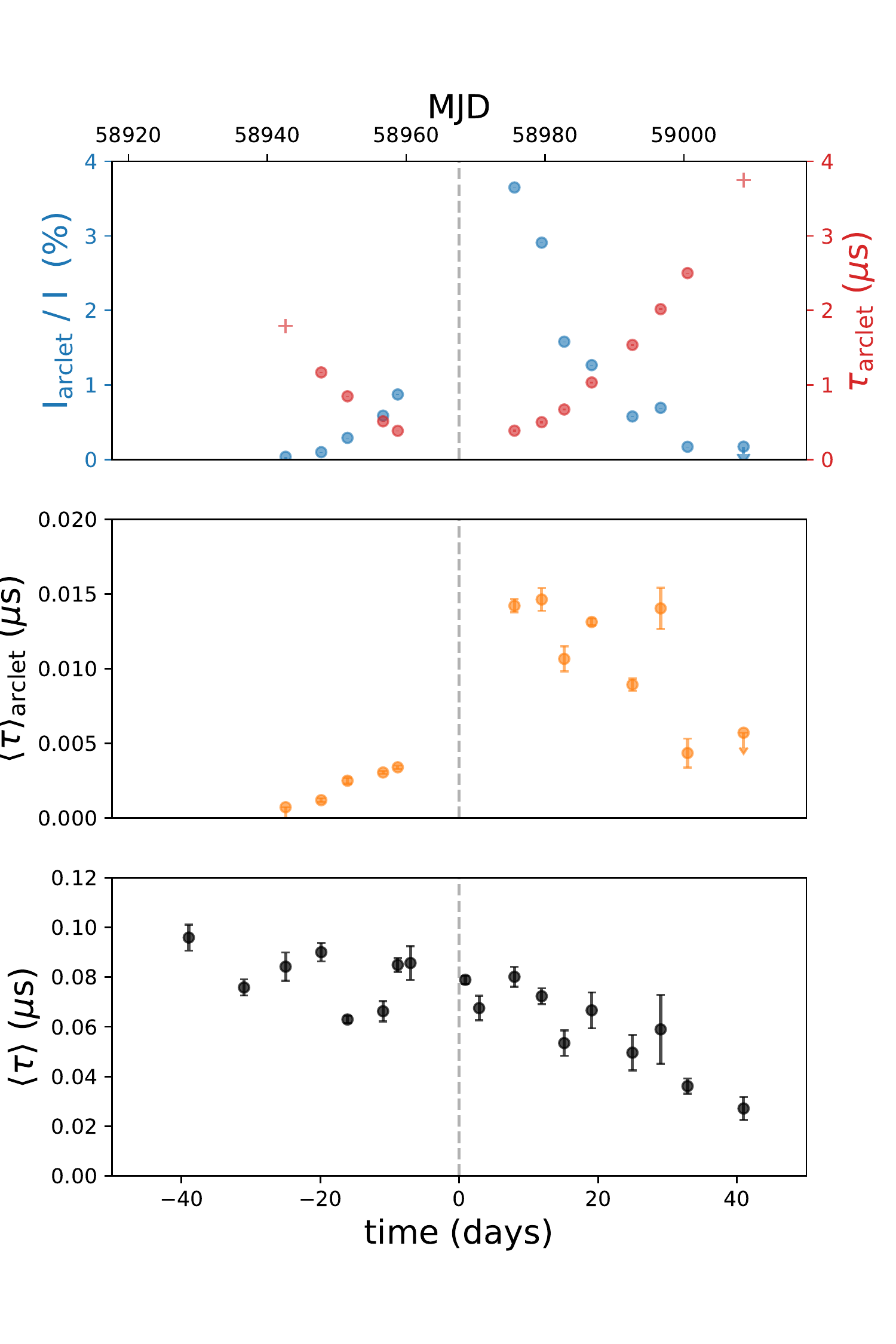}
\caption[Cross correlation]{ Arclet evolution, and time delays from bi-weekly observations with Effelsberg. {\it Top:} Magnification (fraction of total intensity) and time delay of one arclet seen moving through the secondary spectrum. Red crosses are expected positions of the undetected arclet.  {\it Middle:} contribution of the single arclet to the total time delay to the pulsar's signal.  {\it Bottom:}  Estimate of the bulk scattering time inferred from the secondary spectra.
\label{fig:arclet}}
\end{figure}

We investigate the variability of scattering on week to month timescales with the Effelsberg observing campaign described in Section \ref{sec:effelsberg}.
The secondary spectra of all of our observations are shown in Figure \ref{fig:CS2020}.  A clear parabolic arc, with a hint of inverted arclets is seen, and can be seen to clearly move through the secondary spectrum from left to right.
We show three examples of larger, zoomed in secondary spectra in Figure \ref{fig:SecSpecZoom} to emphasise these features.
We estimate the total time delay from the secondary spectrum using the methods of Section \ref{sec:delays}, shown in the bottom panel of Figure \ref{fig:arclet}.  The total time delays are consistent with what was found with LEAP's monitoring, showing steady scattering at around 60--100\,ns, decreasing slightly over two months.   

We also attempt to measure the effect of a single arclet, which is akin to contribution of a pulsar passing a single, compact point of scattering in the ISM, not unlike an echo.  We track the position, and total fractional flux of the arclet seen travelling to the upper-right in the final 7 panels of Figure \ref{fig:CS2020}.  This same feature first appears to be moving towards the origin in panels of MJD 58951-58958, although it is less prominent.  We fit a flux centroid in an ellipse around the arc, to track its motion in $f_{D}$ and $\tau$.  We measure the fraction of the flux in the arc, compared to that integrated over the full parabola, which are plotted in the top panel of Figure \ref{fig:arclet}.  The motion of the arc unsurprisingly traces out a parabola over time,as seen in PSR B0834+06 \citep{hill+05}, and is similar to echoes seen in the Crab pulsar (eg. \citealt{backer+00, lyne+01}).  The strength of the arclet is quite asymmetric about the origin, and at its peak contains $\approx 4\%$ of the total pulsar flux.  The total time shift arising from this arclet can be estimated as $\langle \tau \rangle_{\rm arclet} \approx \tau_{\rm arclet} (I_{\rm arclet} / I)$, and is shown in the middle panel of Figure \ref{fig:arclet}.  The contribution from a single arclet as it passes in front of the pulsar contributes a variable scattering of $\sim 20\,$ns over a 2 month period.

\subsection{Comparison to earlier results}

Previous analysis using the ACF in \citet{levin+16} measures the time delay from scintillation to be $\langle \tau \rangle = 11.7\pm4.9$\,ns, monitoring this pulsar up until October 2013.  Additionally, \citet{shapiro+20} estimate a time delay of $\langle \tau \rangle = 43.6\pm2.3$\,ns in a similar manner.  The frequency channels used in these papers were 1.5625\,MHz wide, a common standard in timing archives, and would have averaged over the fine scintillation structures due to power at high delays.  \citet{keith+13} estimate a scintillation bandwidth of 1.64\,MHz, for which one would infer a time delay of 97\,ns.  This measurement is from before we have data, so we cannot directly compare, but this value is much closer to the order of the time delays we measure.  

\subsection{Effects of uncorrected scattering}

\begin{figure}
\center{}
\includegraphics[width=1.0\columnwidth,trim=0cm 1cm 0cm 0cm, clip=true]{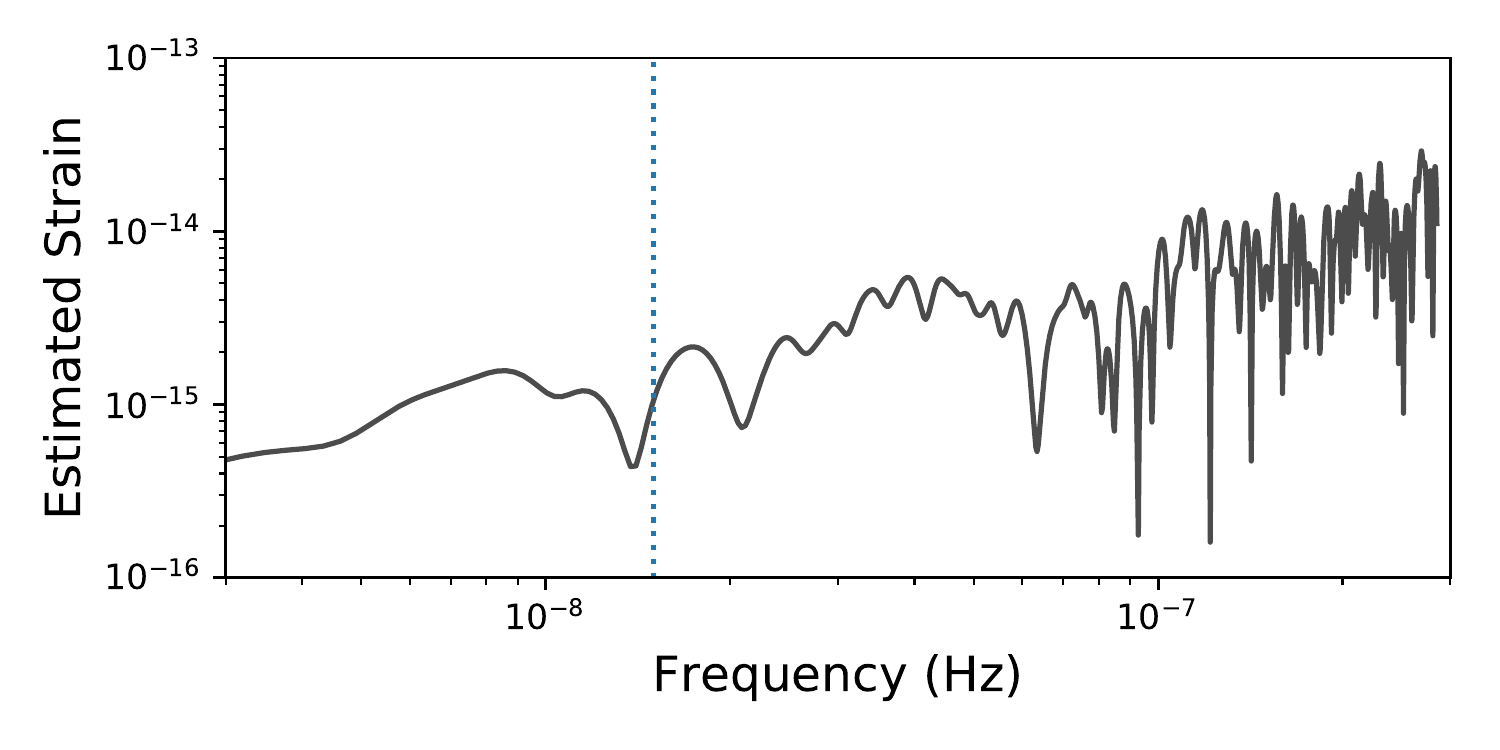}
\caption{ Estimated strain signal which would arise from uncorrected scattering, obtained from a Lomb-scargle periodogram of the bulk time delays $\langle \tau \rangle$ from Figure \ref{fig:scattering}. 
\label{fig:ls}}
\end{figure}

In this section, we aim to make from our measurements a simple estimate of the single-pulsar gravitational wave signal arising from unaccounted scattering.  We stress that this is a conservative upper-limit, as we do not know the extent to which variable scattering is absorbed in red-noise modelling, or in measurements of DM.  The GW strain $h$ at a given periodicity $P$ is related to the amplitude TOA variations $\delta t$ as roughly $h \sim 2\pi \delta t / P$. Since \citet{aggarwal+19} find excess signal at $15\,$nHz, which is $\sim 2.1$\,years, we estimate this using the long-term variable time delays from LEAP shown in Figure \ref{fig:scattering}.  We perform a Lomb-Scargle periodogram on the measured values of $\langle \tau \rangle$, and convert to a measure of $h$ while taking into account the proper normalizations, shown in Figure \ref{fig:ls}.  The measured value at $15\,$nHz is $\sim 10^{-15}$, still an order of magnitude lower than the single pulsar limit of $h = 9.7\times 10^{-15}$ from the EPTA (95\% upper limit, from Table 1 in \citealt{lentati+15}).  As PTA upper limits are improved, scattering variations, if unaccounted for, may begin to limit the timing precision.

\section{Conclusions}
\label{sec:conclusions}

We have measured variable time delays on a PTA millisecond pulsar, using similar methods to those laid out in \citet{hemberger+08}.
The next logical step will be to perform timing with scattering timescales subtracted from TOAs, to see if this improves the timing residuals.  One can apply this approach to study the variable scattering in many PTA pulsars.  With LEAP we can re-reduce the data to whatever time and frequency resolution we like, but regular timing observations would benefit with a second reduction with fine frequency channels at the expense of phase bins.  Going further, methods to obtain the interstellar response directly may be important, including holography \citep{walker+08}, cyclic spectroscopy \citep{demorest11, walker+13, palliyaguru+15, dolch+20}, or directly by using bright giant pulses in special cases \citep{main+17}. Only these methods, in which the interstellar delays are measured directly (as opposed to measuring delay differences, as we do in the secondary spectrum) can retrieve overall delays that are not related to the characteristic timescale of the scattering tail.  New analysis techniques such as the $\theta$-$\theta$ diagram \citep{sprenger+20}, which expresses the secondary spectrum in terms of the angular coordinates on the scattering screen, may be useful as well. This technique can be used to precisely measure arc curvatures, and could be used to efficiently perform holography of 1D screens (Baker et al. in prep.).

Scattering is statistically expected to follow $\sim\lambda^{4}$, yet scattering arising from discrete structures (observed as arclets, or localised clumps of power in secondary spectra) will be localised at a fixed $\tau$ as a function of wavelength.  Observations over a wide frequency range will help to inform the amplitude scaling of arclets, and thus the contribution of discrete arclets to the total scattering time at different frequencies.  In addition, scattering occurs from density gradients in the ISM, so the link between variable DM and scattering should be explored in more detail.  In the case where the DM and scattering variations occur in the same scattering screen, they could potentially be inferred from the other quantity; a predictive model of scattering from DM (or vice versa) would be a great step towards removing these effects from timing observations.  

Knowing the screen distance and orientation, the location of power in the secondary spectrum is predetermined, but the amplitudes are dependent on the physics of scattering and lens models.  In recent years, some predictive models of scintillation properties have been developed  (eg. \citealt{simard+18, gwinn+19}), which can be tested using measurements tracking arclets in time and frequency.

We used annual variations in the arc curvature to determine properties of the scattering screen, while orbital variations were ignored.  Orbital variations can give an additional orbital constraint, such as the inclination (including the ``sense'', \citealt{rickett+14, reardon+19, reardon+20}), and could possibly lead to precise pulsar distances \citep{boyle+12}. Such analysis would be greatly improved with precise, quantitative measurements of the arc curvature. 
Additionally, one could instantaneously measure the scattering screen's properties using the multiple telescopes of LEAP, either using the visibilities \citep{brisken+10} or more simply using the inter-station time delays of the dynamic spectra (\citealt{simard+19}, while the method of combining of visibilities and intensites is outlined in \citealt{simard+19b}).  This is being investigated, and will be the subject of future work.
 
As scattering screens are likely much smaller than the angular separation between pulsars, scattering variations are very unlikely to directly correlate between
pulsars in a way that mimics a Hellings \& Downs curve \citep{hellings+83}.  
But while no direct correlation is expected between pulsars, it is possible that scattering is variable on similar timescales if pulsar proper motions and distances are comparable, and if the screen distance is not at an extreme (ie. not too close to the pulsar or to the Earth).  Several of the EPTA pulsars show variable scintillation arcs, similar to those shown in this paper, and will be subject of future work.  As PTAs become more sensitive, any PTA result relying on a small number of pulsars may need to consider the effects of variable scattering when interpreting the significance of a potential gravitational wave signal.
 

\section*{Acknowledgements}
RAM thanks Daniel Reardon for useful advice on modelling variable arc curvature, and Tim Sprenger for useful comments on the manuscript.

This work is supported by the ERC Advanced Grant ``LEAP", Grant Agreement Number 227947 (PI M.\,Kramer). 
The European Pulsar Timing Array (EPTA) is a collaboration between European Institutes, namely ASTRON (NL), INAF/Osservatorio Astronomico di Cagliari (IT), the Max-Planck-Institut f{\"u}r Radioastronomie (GER), Nan{\c c}ay/Paris Observatory (FRA), The University of Manchester (UK), The University of Birmingham (UK), The University of Cambridge (UK), and The University of Bielefeld (GER), with an aim to provide high-precision pulsar timing to work towards the direct detection of low-frequency gravitational waves.  
 The Effelsberg 100-m telescope is operated by the Max-Planck-Institut f{\"u}r Radioastronomie. 
 Pulsar research at the Jodrell Bank Centre for Astrophysics and the observations using the Lovell Telescope are supported by a consolidated grant from the STFC in the UK. 
The Westerbork Synthesis Radio Telescope is operated by the Netherlands Foundation for Radio Astronomy, ASTRON, with support from NWO. 
The Nan{\c c}ay Radio Observatory is operated by the Paris Observatory, associated with the French Centre National de la Recherche Scientifique. 

The Sardinia Radio Telescope (SRT) is funded by the Department of Universities and Research (MIUR), the Italian Space Agency (ASI), and the Autonomous Region of Sardinia (RAS), and is operated as a National Facility by the National Institute for Astrophysics (INAF). From Mar 2014 - Jan 2016, the SRT data were acquired as part of the Astronomical Validation of the SRT. We thus thank the SRT Astronomical Validation Team, and in particular: M. Burgay, S. Casu, R. Concu, A. Corongiu, E. Egron, N. Iacolina, A. Melis, A. Pellizzoni, and A. Trois. We also thank M. Pilia and A. Possenti for helping with observations. 

 SC acknowledges the support by the ANR Programme d’Investissement d’Avenir (PIA) under the FIRST-TF network (ANR-10-LABX-48-01) project. SC and IC acknowledge financial support from Programme National de Cosmologie and Galaxies (PNCG) and Programme National Hautes Energies (PNHE) funded by CNRS, CEA and CNES, France.
 
 KL is supported by the European Research Council for the ERC Synergy Grant BlackHoleCam under contract no.\ 610058.  K.J.Lee gratefully acknowledges support from National Basic Research Program of China, 973 Program, 2015CB857101 and NSFC 11373011. DP gratefully acknowledges financial support from the research grant “iPeska” (P.I. Andrea Possenti) funded under the INAF national call PRIN-SKA/CTA with Presidential Decree 70/2016. WWZ was supported by the National Natural Science Foundation of China Grant No. 11873067 the CAS-MPG LEGACY project and the Strategic Priority Research Program of the Chinese Academy of Sciences Grant No. XDB23000000.

\section*{Data Availability}
The timing data used this article shall be shared on reasonable request to the corresponding author.

\bibliography{J0613}
\bibliographystyle{mnras}

\label{lastpage}
\end{document}